\documentclass[a4paper,11pt]{article}
\usepackage{jheppub} 

\title{\boldmath Probing Unruh Effect from Enhanced Decoherence}

\author[a,1]{Ran Li\note[1]{Corresponding author.},}

\author[a]{Zhong-Xiao Man,}

\author[b]{Jin Wang}

\affiliation[a]{Department of Physics, Qufu Normal University, Qufu, Shandong 273165, China}

\affiliation[b]{Department of Chemistry, Physics and Astronomy, Stony Brook University, Stony Brook, NY 11794, USA,}

\emailAdd{liran@qfnu.edu.cn}
\emailAdd{zxman@qfnu.edu.cn}
\emailAdd{jin.wang.1@stonybrook.edu}

\abstract{We investigate the decoherence of an Unruh-DeWitt detector coupled to scalar, electromagnetic, and spinor fields in four-dimensional Minkowski spacetime. By employing the Schwinger-Keldysh influence functional formalism, we derive a universal scaling law relating the decoherence rate to the proper acceleration $a$ and the scaling dimension $\Delta$ of the environmental field operator. By analyzing both sharp (top-hat) and smooth Gaussian switching functions, it is shown that the decoherence rate in the asymptotic long-time limit scales as $a^{2\Delta-1}$. This scaling indicates that increasing scaling dimension of the coupling field operators can significantly enhance the decoherence, thereby providing a more sensitive probe of the Unruh effect.}

\begin{document}
\maketitle
\flushbottom

\section{Introduction} 
\label{sec:intro}

The Unruh effect \cite{Unruh:1976db} is one of the most remarkable predictions of quantum field theory in curved spacetime \cite{Birrell:1982ix}. It states that an uniformly accelerated observer in Minkowski spacetime experiences a thermal bath at temperature proportional to the proper acceleration \cite{Crispino:2007eb}. Despite its fundamental importance and its deep connection to Hawking radiation \cite{Davies:1974th}, a direct observation of the Unruh effect in the lab remains extremely challenging due to the extremely small temperature for realistic accelerations. This difficulty has motivated many proposals to search for observable signatures of the Unruh effect. Typical approaches include particle detector models \cite{Unruh:1976db,Dewitt:1979ig}, Berry's phase \cite{Martin-Martinez:2010gnz,Martin-Martinez:2013xer}, quantum simulation \cite{Hu:2018psq,Cheng:2025kyh}, as well as analog gravity \cite{Unruh:1980cg,Garay:1999sk,Leonhardt:2002wej,Fedichev:2003id,Fedichev:2003dj,Philbin:2007ji,Nation:2009xb,Horstmann:2009yh,Steinhauer:2015saa,Zheng:2024drg,Toussaint:2025eap,Deswal:2025cjw,Arya:2024qke}.

An alternative way to probe the Unruh effect is through the observable effects in quantum information theory, such as decoherence. When the detector along an accelerated trajectory interacts with the environmental fields, the vacuum fluctuations can lead to loss of quantum coherence \cite{Xu:2023tdt,Wilson-Gerow:2024ljx}. Unlike detecting the thermal particles directly, decoherence reflects the dynamics of vacuum fluctuations and may provide a more sensitive probe of this kind of acceleration-induced quantum effects \cite{Nesterov:2020exl,Fernandez:2025ijm}.

In this work, we investigate the decoherence of an Unruh–DeWitt detector interacting with the scalar, electromagnetic, and spinor fields in four-dimensional Minkowski spacetime. In our setup, the interaction Hamiltonian commutes with the detector Hamiltonian, which can be treated as a quantum nondemolition type measurement process \cite{Braginsky:1980qv}. The detector undergoes decoherence without dissipation, which can be interpreted as arising from continuous monitoring by vacuum fluctuations of the environmental fields.

Using the Schwinger-Keldysh the influence functional formalism \cite{Schwinger:1960qe,Keldysh:1964ud,Feynman:1963fq,FeynmanHibbs,Calzetta:2008iqa}, we derive the decoherence functional for a detector coupled to the environmental field operators of arbitrary scaling dimension. We further analyze both the sharp (top-hat) and the smooth Gaussian switching functions. Our analysis reveals a universal scaling law relating the decoherence rate to the detector's proper acceleration and the scaling dimension of the environmental field operator. In particular, the decoherence rate scales linearly with acceleration for scalar fields, cubically for electromagnetic fields, and as the fifth power for fermionic fields. The results demonstrate that increasing scaling dimension of the coupling operators can significantly enhance the decoherence induced by the Unruh effect. This also suggests that coupling the detector with higher-dimensional field operator may provide a more sensitive probe of the Unruh effect.

\section{Unruh-DeWitt detectors coupled with the environmental fields}

We consider a two-level Unruh-DeWitt detector as a localized quantum probe moving along a prescribed trajectory $x^\mu(\tau)$ in four dimensional Minkowski spacetime. The Hamiltonian of the detector is given by
\begin{eqnarray}
    H_d=\frac{\Omega}{2} \sigma_z\;,
\end{eqnarray}
where $\Omega$ is the energy gap between the ground state $|g\rangle$ and the excited state $|e\rangle$, and $\sigma_z=|e\rangle\langle e|-|g\rangle\langle g|$ is the Pauli matrix. 

The detector interacts with a quantum field (scalar, electromagnetic, or fermionic) through a quantum nondemolition type coupling, where the detector-field interaction Hamiltonian is given by  
\begin{eqnarray}\label{Int_Ham}
    H_I=\lambda \chi(\tau) \sigma_z \hat{\mathcal{O}}(x(\tau))
\end{eqnarray}
where $\lambda$ is the coupling constant, $\chi(\tau)$ is the switching function controlling the time dependence of the interaction \cite{Higuchi:1993cya}, and $\hat{\mathcal{O}}(x(\tau))$ denotes the coupling field operator along the worldline $x^\mu(\tau)$ of the detector. 

Note that the interaction Hamiltonian in Eq.~\eqref{Int_Ham} is chosen to commute with the free Hamiltonian of the detector. As a result, the populations of the detector's energy eigenstates remain unchanged, and no transitions occur. This type of coupling allows us probe purely the decoherence effects of an Unruh-DeWitt detector. The resulting dynamics corresponds to a quantum nondemolition type measurement process \cite{Braginsky:1980qv}.

In the present work, we will consider the following three cases. For the first case, the detector is coupled with the scalar field. The local field operator is then given by
\begin{eqnarray}
    \hat{\mathcal{O}}(x)=\phi(x)\;.
\end{eqnarray}
This simple model is essentially the scalar monopole model considered in \cite{Wilson-Gerow:2024ljx}. It is shown that the Unruh-Dewitt detector can be mapped to studying the decoherence effects observed for the spatial superposition state in black hole spacetimes \cite{Danielson:2022tdw,Danielson:2022sga,Gralla:2023oya,Biggs:2024dgp,Danielson:2024yru,Kawamoto:2025kfu,Li:2025vcm} and de-Sitter spacetime \cite{Li:2024lfv}. Our study extends beyond this particular case.

For the second case, we consider that the detector is coupled with the electromagnetic field. The field operator $\hat{\mathcal{O}}$ is then defined by 
\begin{eqnarray}
    \hat{\mathcal{O}}(x)=d^i E_i(x)\;,
\end{eqnarray}
where $d^i$ is the detector’s dipole operator and $E_i$ is the electric field at the detector’s position with $i$ denoting the fixed polarization direction of the electric field. 

In the third case, we consider that the detector is coupled with the fermionic field. Unlike scalar or electromagnetic fields, due to the fact that the fermionic operators $\psi(x)$ obey anticommutation relations, the detector cannot couple linearly to $\psi(x)$ in a consistent way. In this case, the interaction must involve a fermion bilinear. For the purpose of studying environment induced decoherence, we consider the feimionic bilinear coupling \cite{Takagi:1986kn,Hummer:2015xaa,Louko:2016ptn,Dubey:2025hwk}, where the field operator $\hat{\mathcal{O}}$ is given by
\begin{eqnarray}
    \hat{\mathcal{O}}(x)=\bar{\psi}(x)\psi(x)\;.
\end{eqnarray}

In the interaction Hamiltonian Eq.\eqref{Int_Ham}, we select the quantum nondemolition type coupling. This type of coupling ensures that the detector’s energy eigenstates are not mixed, leading to pure phase decoherence of off-diagonal density matrix elements without inducing transitions. In the theory of open quantum systems, a standard framework for describing such phenomena is the spin-boson model with a phase-damping coupling\footnote{We thank the anonymous referee for pointing this out.} (see \cite{Hornberger:2008xkz} for a nice review). In Appendix A, we provide a detailed comparison between our model and the spin-boson model discussed in \cite{Hornberger:2008xkz}.

In addition to the detector and interaction parts, the total Hamiltonian of the system also includes the free field contribution. For the three cases discussed above, we take the Hamiltonian $H_{f}$ of the fields to be that of the corresponding free fields.  In the following section, we will employ the Schwinger–Keldysh formalism to study the decoherence effect of the Unruh–DeWitt detector. It is worth emphasizing that the general structure of the Schwinger–Keldysh formalism is independent of the specific form of the free field action. However, the free field actions uniquely determine the propagators of the environmental field operators. These propagators determine the noise kernel appearing in the influence functional and therefore control the scaling behavior of the decoherence rate of the detector.

\section{Decoherence of Unruh-DeWitt detectors}

\subsection{Decoherence functional in Schwinger-Keldysh formulism} 

To study decoherence effect of the detector, we employ the Schwinger-Keldysh (closed time path integral) formalism, which is particularly suited for describing the evolution of reduced density matrices in open quantum systems.

We consider an Unruh-DeWitt detector interacting with a quantum field environment. The total action of the system can be written as
\begin{equation}
S =
S_{d} +
S_{f} +
S_{I},
\end{equation}
where $S_{d}$ denotes the detector action, $S_{d}$ the free field action of the environment, and $S_{I}$ the interaction between the detector and the field.

The density matrix of the total system at the initial time $t_i$ is taken to be the density matrix of a product state 
\begin{equation}\label{Initial_state}
\rho(t_i)=
\rho_{d}(t_i)\otimes
\rho_{\phi}(t_i)\;,
\end{equation}
where $\rho_{d}$ denotes the state of the detector and $\rho_{\phi}$ denotes the state of the environmental field. The density matrix $\rho_{\phi}$ for the environmental field is taken to be in the Minkowski vacuum state, while the initial state of the detector is left arbitrary.

The density matrix of the total system at time $t_f$ is then given by
\begin{equation}
\rho(t_f) = U(t_f,t_i) \,\rho(t_i)\, U^\dagger(t_f,t_i),
\end{equation}
where $U(t_f,t_i)$ is the unitary evolution operator from the initial time $t_i$ to $t_f$.

For a detector coupled to a quantum field, the evolution operator can be expressed as a path integral over all possible histories of the detector $q(\tau)$ and the environmental field $\phi(x)$:
\begin{equation}
U(t_f,t_i) = \int \mathcal{D}q \,\mathcal{D}\phi \; e^{i S[q,\phi]}.
\end{equation}
Here, the environmental fields discussed in the previous section, i.e. the scalar, electromagnetic and fermionic fields, are all denoted by the symbol $\phi$ for simplicity. In fact, $U(t_f,t_i)$ gives the unitary time evolution of a pure state \cite{FeynmanHibbs}.

Substituting the path-integral representation for both $U$ and $U^\dagger$ into the density matrix yields
\begin{align}
\rho(q_f, \phi_f; q_f', \phi_f'; t_f) 
&= \int dq_i \, d\phi_i \, dq_i' \, d\phi_i' \;
\rho(q_i, \phi_i; q_i', \phi_i'; t_i) \nonumber\\
&\quad \times \int_{\left(q_i,\phi_i,q_i',\phi_i'\right)}^{\left(q_f,\phi_f,q_f',\phi_f'\right)} \mathcal{D}q^+ \, \mathcal{D}\phi^+ \,  \mathcal{D}q^- \, \mathcal{D}\phi^- 
\, \exp\Big\{ i \big(S[q^+,\phi^+] - S[q^-,\phi^-]\big) \Big\}.
\end{align}
In the Schwinger-Keldysh formulism \cite{Schwinger:1960qe,Keldysh:1964ud}, the time contour runs forward and backward paths. This introduces two copies of the dynamical variables, where $q^+(\tau)$ and $\phi^+(x)$ denote the forward-time paths associated with $U$, while $q^-(\tau)$ and $\phi^-(x)$ denote the backward-time paths associated with $U^\dagger$. This form of the time-evolved density matrix, which is expressed as a functional integral over all possible histories, naturally allows tracing out the environmental field degrees of freedom to get the reduced density matrix of the detector and to define the influence functional and the decoherence functional for the detector.

The reduced density matrix of the detector at time $t_f$ is obtained by tracing out the environmental field degrees of freedom, which can be expressed as 
\begin{equation}
\rho_{d}(t_f)=
\mathrm{Tr}_{\phi}
\left[
\rho(t_f)
\right]\;.
\end{equation}
Using the path-integral representation of the time-evolution operator and considering the separability of the density matrix of the initial state, the reduced density matrix at time $t_f$ is then given by 
\begin{equation}
\rho_{d}(q_f,q_f',t_f)
=
\int dq_i dq_i' \; \rho_{d}(q_i,q_i',t_i) \int_{\left(q_i,q_i' \right)}^{\left(q_f,q_f'\right)}\mathcal{D}q^+ \mathcal{D}q^- \exp\Big\{ i \big(S_d[q^+] - S_d[q^-]\big) \Big\} \mathcal{F}[q^+,q^-] ,
\end{equation}
where $\mathcal{F}[q^+,q^-]$ is defined as
\begin{eqnarray}
     \mathcal{F}[q^+,q^-]&=&\int d\phi_f d\phi_i d\phi_i'  \int_{\left(\phi_i,\phi_i'\right)}^{\left(\phi_f,\phi_f\right)} \mathcal{D}\phi^+ \mathcal{D}\phi^- \, 
\rho_{\phi}(\phi_i,\phi_i',t_i)\nonumber\\
&&\times  \exp \Big\{ i S_{f}[\phi^+] - i S_{f}[\phi^-] \Big\} 
 \exp \Big\{ i S_I[q^+,\phi^+] - i S_I[q^-,\phi^-] \Big\}.
\end{eqnarray}
The functional $\mathcal{F}[q^+,q^-]$ is called the influence functional, which describes the effect of the environmental field on the time evolution of the reduced density matrix of the detector.

We now consider the linear coupling between the detector and the environmental field \cite{Calzetta:2008iqa}, where the interaction action is given by 
\begin{eqnarray}
    S_I=-\int d\tau q(t) \phi(x(\tau))\;.
\end{eqnarray}
Note that this form of interaction action is consistent with the interaction Hamiltonian given in Eq.\eqref{Int_Ham}. For a Gaussian environment (free field), the path integral over $\phi_\pm$ can be performed exactly. This in turn gives the influence functional as \cite{Calzetta:2008iqa}
\begin{align}
\mathcal{F}[q^+,q^-] = \exp\Big( i S_{\rm IF}[q^+,q^-] \Big),
\end{align}
where the Feynman-Vernon influence action reads
\begin{align}
S_{\rm IF}[q_a,q_r] &= \int d\tau d\tau' \, q_a(\tau) G_R(\tau,\tau') q_r(\tau') 
+ \frac{i}{2} \int d\tau d\tau' \, q_a(\tau) G_H(\tau,\tau') q_a(\tau'), 
\end{align}
with the average variable $q_r$ and the response variable $q_s$:
\begin{align}
    q_r &= \frac{q^+ + q^-}{2}, \quad q_a = q^+ - q^-,
\end{align}
and the retarded Green’s function $G_R$ and the Hadamard function $G_H$ of the environmental field: 
\begin{align}
G_R(\tau,\tau') &= i \theta(\tau-\tau') \langle [\phi(\tau),\phi(\tau')] \rangle_{\rho_\phi},\\
G_H(\tau,\tau') &= \frac{1}{2} \langle \{ \phi(\tau),\phi(\tau') \} \rangle_{\rho_\phi}.
\end{align}
Here, the vacuum expectation values are taken with respect to the Minkowski vacuum of the environmental fields.

It is clear that the real part of the Feynman-Vernon influence action only contributes a phase and governs dissipative dynamics and the imaginary part produces the real exponential damping factor of the influence functional. It is the imaginary part that suppresses interference between different system histories and leads to the decay of off-diagonal elements of the reduced density matrix. Therefore, the decoherence functional is only determined by the imaginary part of the Feynman-Vernon influence action.

Based on the above discussion, the decoherence functional is defined as
\begin{equation}\label{Dec_func}
\Gamma[q^+,q^-]
=
\frac{\lambda^2}{2}
\int d\tau d\tau'
\,\chi(\tau)\chi(\tau')
\left(q^+(\tau)-q^-(\tau)\right)
\left(q^+(\tau')-q^-(\tau')\right)
G_H(x(\tau),x(\tau')),
\end{equation}
where
\begin{equation}
G_H(x,x')
=
\frac{1}{2}
\langle \{\hat{\mathcal{O}}(x),\hat{\mathcal{O}}(x')\} \rangle
\end{equation}
is the Hadamard correlation function of the environmental field. Here, we have restored the explicit expression of the interaction Hamiltonian given in Eq.\eqref{Int_Ham}. This expression for the decoherence functional shows that the decoherence effect is governed by the symmetrized two-point correlation function of the environmental field evaluated along the detector trajectory.

This formalism applies for scalar, electromagnetic, and fermionic fields. For fermionic fields, the coupling is bilinear, e.g., $\hat{\mathcal{O}}[\psi] = \bar\psi \psi$, and the trace over the Grassmann fields yields an influence functional of the same exponential form, with the propagators replaced by the fermionic Wightman functions.

\subsection{Decoherence functional of Unruh-DeWitt detectors}

We now apply the decoherence functional obtained from the Schwinger-Keldysh formlism to the Unruh-DeWitt detectors.

For a two-level system with coupling through the Pauli matrix $\sigma_z$, the off-diagonal elements $\rho_d^{+-}$ and $\rho_d^{-+}$ of the reduced density matrix correspond to
\begin{equation}
\sigma_z^+ = +1, 
\qquad 
\sigma_z^- = -1.
\end{equation}
The forward and backward branches of the Schwinger–Keldysh contour
correspond to different eigenstates of $\sigma_z$. This yields 
\begin{equation}
\sigma_z^+ - \sigma_z^- = 2 .
\end{equation}
Substituting this into Eq.\eqref{Dec_func} gives the final form of the decoherence functional 
\begin{equation}\label{Dec_GH}
\Gamma
=
2\lambda^2
\int d\tau d\tau'
\,\chi(\tau)\chi(\tau')
G_H(x(\tau),x(\tau')) .
\end{equation}
This functional characterizes the decoherence effect of the Unruh-Dewitt detector because the off-diagonal elements of the reduced density matrix for the Unruh-Dewitt detector decay as
\begin{equation}\label{off_dia_rho}
\rho_d^{+-}(t_f)
\propto
\rho_d^{+-}(t_i)\exp(-\Gamma).
\end{equation}

We consider the Unruh-Dewitt detector moving along a uniformly accelerated trajectory with proper acceleration $a$ in the $x$ direction with respect to an inertial reference frame in four-dimensional Minkowski spacetime. The trajectory can be parametrized by the detector's proper time $\tau$ as
\begin{align}
t(\tau) &= a^{-1}\sinh(a\tau),\\
x(\tau) &= a^{-1}\cosh(a\tau),\\
y(\tau) &= 0, \\
z(\tau) &= 0 .\\
\end{align}
This trajectory satisfies
\begin{equation}
x^2 - t^2 = \frac{1}{a^2},
\end{equation}
which corresponds to a hyperbola in Minkowski spacetime. The parameter $a$ represents the constant proper acceleration experienced by the detector. Along this trajectory, the proper time $\tau$ is related to the Minkowski coordinates $(t,x,y,z)$ through the above parametrization, and the detector remains confined to the Rindler wedge $x>|t|$.

For stationary environments, the Hadamard correlation function depends only on the proper time difference
\begin{equation}
\Delta\tau = \tau-\tau'.
\end{equation}
The decoherence functional becomes
\begin{equation}
\Gamma =
2\lambda^2
\int d\tau
\int d\Delta\tau
\,
\chi(\tau)\chi(\tau-\Delta\tau)
G_H(\Delta\tau).
\end{equation}

To express this in the frequency domain, we introduce the Fourier transform of the switching function
\begin{equation}
\tilde{\chi}(\omega)
=
\int_{-\infty}^{\infty} d\tau \,
e^{i\omega\tau}\chi(\tau),
\end{equation}
and the Fourier transform of the Hadamard function
\begin{equation}
\tilde{G}_H(\omega)
=
\int_{-\infty}^{\infty} d\Delta\tau \,
e^{i\omega\Delta\tau}
G_H(\Delta\tau).
\end{equation}
Using the identity
\begin{equation}
G_H(\tau-\tau')
=
\int_{-\infty}^{\infty}
\frac{d\omega}{2\pi}
e^{-i\omega(\tau-\tau')}
\tilde{G}_H(\omega),
\end{equation}
the decoherence functional becomes
\begin{align}
\Gamma
&=
2\lambda^2
\int d\tau d\tau'
\chi(\tau)\chi(\tau')
\int \frac{d\omega}{2\pi}
e^{-i\omega(\tau-\tau')}
\tilde{G}_H(\omega) .
\end{align}
Rearranging the integrals gives the decoherence functional in the frequency domain
\begin{align}\label{Dec_func_freq}
\Gamma
&=
2\lambda^2
\int_{-\infty}^{\infty}
\frac{d\omega}{2\pi}
\tilde{G}_H(\omega)
\left|
\tilde{\chi}(\omega)
\right|^2,
\end{align}
which shows that the decoherence is determined by the overlap between the detector switching spectrum and the environmental noise spectrum.

In the following sections, we will explicitly evaluate the decoherence functional obtained from the Schwinger-Keldysh formalism for different types of environmental fields. Our goal is to determine the long-time decoherence rate of the detector and its dependence on the acceleration. This needs to specify the switching function that controlling the coupling between the detector and the environmental fields.

\section{Decoherence rates for top-hat switching function}

\subsection{Top-hat switching function}

We first consider the simplest finite-time coupling, where the switching function is given by 
\begin{equation}\label{Top_hat_switching}
\chi(\tau)=
\begin{cases}
1, & -\frac{T}{2} < \tau < \frac{T}{2},\\
0, & \text{otherwise}.
\end{cases}
\end{equation}
The kind of top-hat switching function corresponds to an idealized situation in which the interaction between the detector and the environmental field is turned on and off instantaneously. It is expected that the decoherence rate becomes insensitive to the detailed form of the switching function in the long interaction time limit. In this regime the top-hat switching provides a convenient analytical approximation that allows one to extract the universal scaling behavior of the decoherence rate.

The Fourier transform of the switching function is given by
\begin{equation}
\tilde{\chi}(\omega)
=
\int_{-\infty}^{\infty} d\tau\, e^{i\omega\tau}\chi(\tau).
\end{equation}
For the top-hat switching function presented in Eq.\eqref{Top_hat_switching}, we obtain
\begin{equation}
\tilde{\chi}(\omega)
=
\int_{-T/2}^{T/2} d\tau\, e^{i\omega\tau}
=
\frac{2\sin(\omega T/2)}{\omega}.
\end{equation}
Then one can get
\begin{equation}
|\tilde{\chi}(\omega)|^2
=
\frac{4\sin^2(\omega T/2)}{\omega^2}.
\end{equation}
Substituting the switching function spectrum into the decoherence functional presented in Eq.\eqref{Dec_func_freq} gives
\begin{equation}
\Gamma(T)
=2
\lambda^2
\int_{-\infty}^{\infty}
\frac{d\omega}{2\pi}
\frac{4\sin^2(\omega T/2)}{\omega^2}
\tilde{G}_H(\omega).
\end{equation}

In the limit of the long interaction time, i.e. $T \gg a^{-1}$, we use the approximation 
\begin{equation}
\frac{\sin^2(\omega T/2)}{\omega^2}
\approx
\frac{\pi T}{2}\delta(\omega),
\end{equation}
which leads to
\begin{equation}
\Gamma(T)
\approx
2\lambda^2 T\,\tilde{G}_H(0).
\end{equation}
The decoherence rate is therefore given by 
\begin{equation}
\gamma
=
\frac{\Gamma(T)}{T}
=
2\lambda^2 \tilde{G}_H(0).
\end{equation}

From the above formula for the decoherence rate, one can see that the decoherence of the Unruh-DeWitt detector with a pure dephasing coupling is determined by low-frequency environmental fluctuations. In the aspect, the environmental field acts as a quantum thermal bath, and its vacuum fluctuations along the detector trajectory determine the decoherence rate. For an accelerated detector, the spectrum $\tilde{G}_H(0)$ acquires a thermal structure due to the Unruh effect. Therefore, it is expected that the decoherence rate depends on the proper acceleration $a$ and the decoherence of the detector can provide a probe to the Unruh effect.

\subsection{Decoherence rate in the scalar field environment}

We first consider a massless scalar field in Minkowski spacetime and evaluate the Hadamard correlation functions along the worldline of a uniformly accelerated detector. For a massless scalar field in four-dimensional Minkowski spacetime, the Wightman function is given by \cite{Birrell:1982ix}
\begin{equation}\label{scalar_function}
G^{+}(x,x')=\langle0|\phi(\tau)\phi(\tau')|0\rangle
=
-\frac{1}{4\pi^2}
\frac{1}{(t-t'-i\epsilon)^2-|\mathbf{x}-\mathbf{x'}|^2},
\end{equation}
where $|0\rangle$ denotes the vacuum state with respect to the inertial observer in the Minkowski spacetime. It can be seen that the Wightman function depends on the invariant interval of the Minkowski spacetime. The invariant interval between two points on the accelerated trajectory is given by
\begin{equation}
-(t-t'-i\epsilon)^2+(x-x')^2
=-
\frac{4}{a^2}
\sinh^2
\left(
\frac{a(\tau-\tau'-i\epsilon)}{2}
\right).
\end{equation}
By substituting this into Eq.\eqref{scalar_function}, one can get the Wightman function along the accelerated trajectory
\begin{equation}
G^+(\Delta\tau)
=-\frac{a^2}{16\pi^2}
\frac{1}{\sinh^2\left(\frac{a\Delta\tau-i\epsilon}{2}\right)}.
\end{equation}

The Fourier transform of the Wightman function is defined as
\begin{equation}
\tilde G^{+}(\omega)
=
\int_{-\infty}^{\infty}
d\Delta\tau \,
e^{i\omega\Delta\tau}
G^{+}(\Delta\tau).
\end{equation}
Using contour integration, one obtains
\begin{equation}
\tilde G^{+}(\omega)
=
\frac{\omega}{2\pi}
\frac{1}{1-e^{-2\pi\omega/a}}.
\end{equation}
This spectrum satisfies the KMS thermal relation \cite{Kubo:1957mj,Martin:1959jp}
\begin{equation}
\tilde G^{+}(-\omega)
=
e^{-2\pi\omega/a}
\tilde G^{+}(\omega),
\end{equation}
corresponding to a thermal bath with the Unruh temperature
\begin{equation}
T=\frac{a}{2\pi}.
\end{equation}
The Unruh effect reflects the fact that the Minkowski vacuum satisfies the KMS condition when restricted to accelerated observers.

The Hadamard correlation function is defined as the symmetrized correlation function
\begin{equation}
G_H(\Delta\tau)
=
\frac12
\langle
\{\phi(\tau),\phi(\tau')\}
\rangle
=
\frac12\left(G^{+}(\Delta\tau)+G^{+}(-\Delta\tau)\right).
\end{equation}
The Fourier transform of the Hadamard correlation function is given by
\begin{equation}
\tilde G_H(\omega)
=\frac12\left(
\tilde G^{+}(\omega)
+
\tilde G^{+}(-\omega)\right).
\end{equation}
Substituting the thermal spectrum yields
\begin{align}
\tilde G_H(\omega)
&=
\frac{\omega}{4\pi}
\left[
\frac{1}{1-e^{-2\pi\omega/a}}
+
\frac{1}{e^{2\pi\omega/a}-1}
\right].
\end{align}
After simplification one obtains
\begin{equation}
\tilde G_H(\omega)
=
\frac{\omega}{4\pi}
\coth\left(\frac{\pi\omega}{a}\right).
\end{equation}

In the low-frequency limit $\omega\ll a$, the Hadamard spectrum behaves as
\begin{equation}
\tilde G_H(\omega\rightarrow 0)
=
\frac{a}{4\pi^2}.
\end{equation}
This constant low-frequency limit is responsible for the decoherence rate of the accelerated detector. Substituting into the decoherence rate expression gives
\begin{equation}
\gamma_s =\lambda^2
\frac{a}{2\pi}.
\end{equation}

Thus, the scalar field environment induces a constant decoherence rate proportional to the proper acceleration of the Unruh–DeWitt detector, or equivalently, to the Unruh temperature. This suggests that the Unruh effect can be probed by the decoherence of the detector. Nevertheless, observing such decoherence in the lab is no less challenging than directly detecting the Unruh effect because the decoherence rate depends on the Unruh temperature linearly. In the following subsections, we show that coupling the detector to environmental field operators with higher scaling dimensions can significantly enhance the sensitivity of decoherence induced by the Unruh effect.

For a uniformly accelerated detector in Minkowski spacetime, the Unruh effect implies that the detector perceives a thermal bath with temperature proportional to its proper acceleration. As a result, the detector behaves effectively as if it were immersed in a thermal environment. As discussed in Appendix A, our model is equivalent to the spin-boson model considered in \cite{Hornberger:2008xkz}. In that context, it is known that in the late-time limit the decoherence rate scales linearly with the environmental temperature. This provides a natural interpretation of our result in terms of the Unruh temperature. However, we will go beyond the scalar-field case and extend our analysis to situations where the detector is coupled to electromagnetic and fermionic environments in the following sections. 

\subsection{Decoherence rate in the electromagnetic field environment}

Now we consider a uniformly accelerated Unruh--DeWitt detector interacting with the electromagnetic field in four-dimensional Minkowski spacetime.

In the Feynman gauge, the Wightman function of the electromagnetic vector potential is given by \cite{Birrell:1982ix}
\begin{equation}\label{vector_prop}
\langle 0|A_\mu(x)A_\nu(x')|0\rangle
=
\eta_{\mu\nu} G^{+}(x,x'),
\end{equation}
where $G^{+}(x,x')$ is the Wightman function of the scalar field given in Eq.\eqref{scalar_function}. Here, we have adopted the Feynman gauge, which is defined as $\partial_\mu A^\mu=0$. This makes the correlation presented in a particularly simple form.

For the electromagnetic field strength tensor defined as $F_{\mu\nu}=\partial_\mu A_\nu-\partial_\nu A_\mu$, the electric field operator is given by 
\begin{equation}
E_i = F_{0i} = \partial_0 A_i - \partial_i A_0 .
\end{equation}
Then two-point correlation function of the electric field can be expressed in terms of derivatives of the vector propagator
\begin{align}
\langle 0|E_i(x)E_j(x')|0\rangle
&=
\langle (\partial_0 A_i-\partial_i A_0)
(\partial'_0 A_j-\partial'_j A_0)\rangle .
\end{align}
Expanding the derivatives gives
\begin{align}
\langle E_i(x)E_j(x')\rangle
&=
\partial_0\partial'_0
\langle A_i A_j\rangle
-
\partial_0\partial'_j
\langle A_i A_0\rangle
-
\partial_i\partial'_0
\langle A_0 A_j\rangle
+
\partial_i\partial'_j
\langle A_0 A_0\rangle .
\end{align}
Using Eq.\eqref{vector_prop}, we obtain
\begin{equation}
\langle 0|E_i(x)E_j(x')|0\rangle
=
\left(
\delta_{ij}\partial_0\partial'_0
-
\partial_i\partial'_j
\right)
G^{+}(x,x').
\end{equation}
Substituting the scalar propagator gives
\begin{equation}
\langle 0|E_i(x)E_j(x')|0\rangle
=
\frac{1}{4\pi^2}
\left(
\delta_{ij}\partial_0\partial'_0
-
\partial_i\partial'_j
\right)
\frac{1}{-(t-t'-i\epsilon)^2+|\mathbf{x}-\mathbf{x'}|^2}.
\end{equation}
For a detector undergoing uniform acceleration $a$ in the $x$ direction, $G^+(x,x')$ depends only on the invariant interval
\begin{equation}
\sigma=-(t-t')^2+|\mathbf{x}-\mathbf{x'}|^2 .
\end{equation}
By noting that 
\begin{eqnarray}
    \partial_\mu \sigma=2(x_\mu-x_\mu')\;,
\end{eqnarray}
one can get the following expression for the correlation function of the electric field after some algebras 
\begin{align}
\langle 0|E_i(x)E_j(x')|0\rangle
&=-\frac{1}{4\pi^2}
\left[
\frac{4\delta_{ij}}{\sigma^2}
+
\frac{8}{\sigma^3}
\left(
\delta_{ij}(t-t')^2
-
(x_i-x_i')(x_j-x_j')
\right)
\right].
\end{align}

We consider the polarization direction of the electric dipole is along the $x$-direction, i.e. $d^\mu=d \hat{x}$. Substituting the trajectory of the accelerating detector into the correlation function, the Hadamard function can be derived as
\begin{equation}
G_{H}(\Delta \tau)= d^2 \langle 0|E_x(x)E_x(x')|0\rangle
= \frac{d^2 a^4}{16 \pi^2 } \frac{1}{\sinh^4\left(\frac{a \Delta \tau}{2}\right)}.
\end{equation}
The Fourier transformation is given by
\begin{eqnarray}
    \tilde G_H(\omega)
&=\frac{d^2}{3\pi}\omega(\omega^2+a^2)
\frac{1}{1-e^{-2\pi\omega/a}}.
\end{eqnarray}

In this case, the low-frequency limit $\omega\ll a$ of the Hadamard spectrum gives 
\begin{equation}
\tilde G_H(\omega\rightarrow 0)
=
\frac{d^2a^3}{6\pi^2}.
\end{equation}
This constant low-frequency limit is then responsible for the decoherence rate of the accelerated detector in the electromagnetic environment. Substituting into the decoherence rate expression gives
\begin{equation}
\gamma_e =\lambda^2 d^2
\frac{a^3}{3\pi^2}.
\end{equation}
Consequently, the decoherence rate induced by the electromagnetic field bath scales with the cube of the detector’s proper acceleration, or equivalently, with the cube of the Unruh temperature. This cubic dependence makes the decoherence rate in the electromagnetic environment much more sensitive to probe the Unruh temperature.

Compared with the linear scaling for the scalar field, the decoherence rate scales cubically with the proper acceleration for the electromagnetic field. This difference arises from the derivative structure of the electromagnetic field operator. It is generally expected that the field operator with higher scaling dimension involves more derivatives. This in turn increases the sensitivity of the decoherence rate to acceleration. In other words, for the field operators with higher scaling dimensions, the Unruh effect produces stronger low frequency noise, which can in principle enhance the detectability of the decoherence induced by the Unruh effect.

\subsection{Decoherence rate in the ferminionc field}

Now, we consider the detector coupled with the environmental ferminionic field. For a massless Dirac field in Minkowski spacetime, the positive-frequency Wightman function is defined as
\begin{equation}
S^+(x,x')_{\alpha\beta} \equiv \langle 0 | \psi_\alpha(x) \bar{\psi}_\beta(x') | 0 \rangle.
\end{equation}
It satisfies the Dirac equation and, in four dimensions, can be written in terms of the scalar propagator as \cite{Birrell:1982ix}
\begin{equation}\label{Fermion_wightman}
S^+(x,x') = i \gamma^\mu \partial_\mu G^+(x,x'),
\end{equation}
where $\gamma^\mu$ is the gamma matrices and $G^+(x,x')$ is the massless scalar Wightman function presented in Eq.\eqref{scalar_function}.

In our model, the detector couples to the fermion bilinear $\bar{\psi}\psi$, so the relevant two-point correlation function is
\begin{equation}
G(x,x') = \langle 0 | \bar{\psi}(x)\psi(x) \bar{\psi}(x')\psi(x') | 0 \rangle.
\end{equation}
Using Wick's theorem for free fermions, this can be expressed as \cite{Louko:2016ptn}
\begin{equation}
G(x,x') = - \textrm{Tr} \big[ S^+(x,x') S^+(x',x) \big],
\end{equation}
Substituting the fermionic Wightman function into the above equation, and utilizing the convention $\textrm{Tr}(\gamma^\mu\gamma^\nu)=4\eta^{\mu\nu}$, one can get
\begin{eqnarray}
   G(x,x') =4 \partial_\mu G^+(x,x') \partial^\mu G^+(x,x').
\end{eqnarray}
Using the form of the scalar propagator, one can get
\begin{eqnarray}
    G(x,x') = \frac{1}{\pi^4} \frac{1}{\left[-(t-t'-i\epsilon)^2+|\mathbf{x}-\mathbf{x'}|^2\right]^3}.
\end{eqnarray}

For a detector uniformly accelerated along the $x$-direction with proper acceleration $a$, the correlation function in the fermionic environment is then given by  
\begin{eqnarray}
    G(\Delta \tau) =- \frac{a^6}{64\pi^4} \frac{1}{\sinh^6\left( \frac{a(\tau-\tau'-i\epsilon)}{2}\right)}.
\end{eqnarray}
The symmetrized Hadamard function is then given by 
\begin{equation}
G_H(\Delta\tau) = \frac12 \langle \{ \bar\psi\psi(\tau), \bar\psi\psi(\tau') \} \rangle
= \frac12 \left[ G(\Delta\tau) + G(-\Delta\tau) \right].
\end{equation}
Its Fourier transform defines the noise spectrum:
\begin{equation}
\tilde G_H(\omega) = \int_{-\infty}^{\infty} d\Delta\tau \, e^{i\omega \Delta\tau} G_H(\Delta\tau)
= \frac{1}{120\pi^3} \omega \left(\omega^2+a^2\right)\left(\omega^2+4a^2\right)\coth\left(\frac{\pi\omega}{a}\right).
\end{equation}
In this case, the low-frequency limit $\omega\ll a$ of the Hadamard spectrum gives 
\begin{equation}
\tilde G_H(\omega\rightarrow 0)
=
\frac{a^5}{30\pi^4}.
\end{equation}
This constant low-frequency noise is responsible for the decoherence rate of the accelerated detector in the fermionic environment. Substituting into the decoherence rate expression gives
\begin{equation}
\gamma_f =\lambda^2 
\frac{a^5}{15\pi^4}.
\end{equation}

From this result, one can see that the decoherence rate of the Unruh-Dewitt detector associated with fermionic fields environment increases much more rapidly with acceleration. Scalar, electromagnetic, and fermionic fields lead to qualitatively different detector responses to acceleration. Among them, fermionic fields produce the strongest dependence on acceleration, leading to a more pronounced signal. For this reason, detectors coupled to fermionic fields can, in principle, provide a more sensitive probe of the Unruh effect than detectors interacting with scalar or electromagnetic fields.

\subsection{The dependence of decoherence rate on the scaling dimension of the operator}

In this subsection, we will interpret that the scaling behavior follows naturally from the scaling dimension $\Delta$ of the corresponding fields coupled to the detector. It is known that the scaling dimensions of the environmental fields are 
\begin{itemize}
\item Scalar field \(\phi\) : $\Delta=1$;
\item Electromagnetic field \(E_i\) : $\Delta=2$;
\item Fermion bilinear \(\bar\psi \psi\) : $\Delta=3$.
\end{itemize}

For a quantum field operator with scaling dimension $\Delta$, the Wightman function behaves generally as
\begin{equation}
G^+(x,x')
\sim
\frac{1}{\sigma^{2\Delta}}.
\end{equation}
Along the accelerated trajectory this gives
\begin{equation}
G^+(\Delta\tau)
\sim
\left[
\sinh\left(\frac{a\Delta\tau}{2}\right)
\right]^{-2\Delta}.
\end{equation}
Using the recurrence relation 
\begin{eqnarray}
    \sinh^{-2n-2}x=\frac{1}{2n(2n+1)}\left(\frac{d^2}{dx^2}-4n^2\right)\sinh^{-2n}x\;,
\end{eqnarray}
the corresponding spectral density of the correlation function behaves as
\begin{equation}
\tilde G_H(\omega)
\sim
\omega \prod_{n=1}^{\Delta-1}(\omega^2+n^2a^2)
\coth\!\left(\frac{\pi\omega}{a}\right).
\end{equation}
In the low-frequency limit one finds
\begin{equation}
\tilde G_H(0)
\propto
a^{2\Delta-1}.
\end{equation}
Therefore the decoherence rate of Unruh-Dewitt detector in different environments scales as
\begin{equation}
\gamma
\propto
\lambda^2 a^{2\Delta-1}.
\end{equation}

It is shown that the decoherence rate is determined by the scaling dimensions of the coupling operators. This result is also fully consistent with the results obtained in the previous cases. For a uniformly accelerated detector, the scaling of the decoherence rate with acceleration in different environments is:
\begin{align}
\text{Scalar field: } & \gamma_s \propto a, \\
\text{Electromagnetic field: } & \gamma_{e} \propto a^3, \\
\text{Fermion field: } & \gamma_f \propto a^5.
\end{align}

Generally, for a accelerating detector coupled to an operator with scaling dimension $\Delta$, the decoherence rate follows the universal scaling law $\gamma \propto a^{2\Delta-1}$. This implies that detectors interacting with the operators with higher scaling dimension experience substantially stronger decoherence. 

A major difficulty in observing the Unruh effect lies in the extremely large acceleration required to produce a measurable temperature, since the Unruh temperature $T=\frac{a}{2\pi}$ is proportional to Planck’s constant when physical units are restored. Reaching temperatures of order $1\,\mathrm{K}$ would require accelerations on the order of $10^{20}\,\mathrm{m/s^2}$, which is far beyond the present experimental capabilities. However, the decoherence framework can significantly amplify the observable signal because the decoherence rate grows with higher powers of the acceleration depending on the operator dimension. Combined with modern quantum platforms capable of engineering controlled couplings, these features make experimental investigations of Unruh-induced decoherence a plausible possibility. This can in turn provide an alternative way to probe the Unruh effect. 

\section{Decoherence under Gaussian switching}

In realistic physical systems, the coupling between a detector and its environment cannot change discontinuously. Therefore the top-hat switching function should be regarded as an idealized model that captures the main features of a finite-time interaction while ignoring the detailed behavior at the switching moments. In more realistic situations one may employ smooth switching profiles, such as Gaussian functions.

We now consider a Gaussian switching function for the detector, which is explicitly given by 
\begin{equation}
\chi(\tau)=\chi_0
\exp\!\left(-\frac{\tau^2}{2\sigma^2}\right). 
\end{equation}
Here, $\chi_0$ determines the effective path separation and $\sigma$ characterizes the interaction time scale.

With the Gaussian switching function, the decoherence functional is then given by 
\begin{eqnarray}
\Gamma
&=&2
\lambda^2
\int d\tau d\tau'
\,
\chi(\tau)\chi(\tau')
G_H(\tau-\tau')\nonumber\\
&=&2
\lambda^2 \chi_0^2
\int d\tau d\tau'
\,
e^{-\frac{\tau^2+\tau'^2}{2\sigma^2}}
G_H(\tau-\tau').
\end{eqnarray}

Introducing the variables
\begin{equation}
u=\tau-\tau', \qquad v=\tau+\tau',
\end{equation}
one finds
\begin{equation}
\tau^2+\tau'^2=\frac{u^2+v^2}{2}, \qquad
d\tau d\tau'=\frac12\,du\,dv .
\end{equation}
Employing the new variables, the decoherence functional becomes
\begin{equation}
\Gamma
=
\lambda^2\chi_0^2
\int_{-\infty}^{\infty} dv\,
e^{-v^2/(4\sigma^2)}
\int_{-\infty}^{\infty} du\,
e^{-u^2/(4\sigma^2)}G_H(u).
\end{equation}

It is easy to see that the $v$ integral can be performed directly. Using the following identity 
\begin{equation}
\int_{-\infty}^{\infty} dv\,e^{-v^2/(4\sigma^2)}
=
\sqrt{4\pi}\sigma ,
\end{equation}
the decoherence functional can be further written in the form of 
\begin{equation}
\Gamma
=
\sqrt{4\pi}\lambda^2\chi_0^2\sigma
\int_{-\infty}^{\infty} du\,
e^{-u^2/(4\sigma^2)}G_H(u).
\end{equation}

For an operator $\hat{\mathcal{O}}$ of scaling dimension $\Delta$ along a uniformly accelerated trajectory, the Hadamard correlation function behaves as 
\begin{equation}
G_H(u)
=
\frac{C_\Delta a^{2\Delta}}
{\sinh^{2\Delta}\!\left[\frac{a}{2}(u-i\epsilon)\right]} ,
\end{equation}
where $C_\Delta$ is the normalization constant of the Hadamard function for an operator of scaling dimension $\Delta$. It ensures that the two-point correlation function has the correct units and overall magnitude. 

Using the exponential series representation
\begin{equation}
\frac{1}{\sinh^{2\Delta}x}
=
2^{2\Delta}
\sum_{n=0}^{\infty}
\frac{\Gamma(n+2\Delta)}{n!\,\Gamma(2\Delta)}
e^{-(2n+2\Delta)x},
\end{equation}
the integral of the decoherence functional reduces to
\begin{equation}
\Gamma
=\sqrt{4\pi}
\lambda^2\chi_0^2 C_\Delta a^{2\Delta}\sigma
2^{2\Delta+1}
\sum_{n=0}^{\infty}
\frac{\Gamma(n+2\Delta)}{n!\,\Gamma(2\Delta)}
I_n ,
\end{equation}
where
\begin{equation}
I_n=
\int_{0}^{\infty}du\,
e^{-u^2/(4\sigma^2)}
e^{-a(n+\Delta)u}.
\end{equation}
This Gaussian integral with a linear exponential term admits a closed representation in terms of the complementary error function $\mathrm{erfc}(x)$. The result is given by 
\begin{equation}
I_n=\sigma \sqrt{\pi} \; e^{a^2 \sigma^2 (n+\Delta)^2} \;
\mathrm{erfc}\big[a \sigma (n+\Delta)\big] .
\end{equation}
This equation, together with the expression for $\Gamma$, constitutes a closed-form expression for the decoherence functional under Gaussian switching.

In the long-interaction limit $a\sigma \gg 1$, the complementary error function admits the asymptotic expansion
\begin{equation}
\mathrm{erfc}(x) \approx \frac{e^{-x^2}}{x \sqrt{\pi}}, \qquad x \gg 1.
\end{equation}
Applying this to the $n$-sum, the dominant contribution comes from $n=0$, yielding
\begin{equation}
\Gamma \approx \sqrt{4\pi} \sigma\lambda^2 \chi_0^2  \, C_\Delta \, a^{2\Delta-1} 2^{2\Delta+1} \sum_{n=0}^{\infty}\frac{\Gamma(n+2\Delta)}{n!\,\Gamma(2\Delta)}\frac{1}{n+\Delta} .
\end{equation}

Thus, in the long-time limit, the decoherence functional saturates to a finite value that scales as a power of the proper acceleration determined by the operator's scaling dimension. This reproduces the universal scaling law for the decoherence rate by taking the $\sigma$ as the interaction time scale 
\begin{equation}\label{scaling_law}
\gamma \equiv \frac{\Gamma}{\sigma} \propto a^{2\Delta-1},
\end{equation}
which clearly shows that the operators with higher scaling dimensions experience stronger Unruh-induced decoherence.  

The scaling law derived above for $D=4$ dimensions can be straightforwardly generalized to arbitrary spacetime dimensions. The key observation is that along the worldline of an uniformly accelerated detector, the two-point function of an operator with scaling dimension $\Delta$ is independent of the spacetime dimension. By applying a similar analysis of decoherence, as shown previously for both top-hat and Gaussian switching, one can obtain the result that the scaling law for the decoherence rate in $D$-dimensional spacetime is also given by Eq.\eqref{scaling_law}. However, the spacetime dimension $D$ does affect the scaling dimension $\Delta$ of the operator.

The decoherence functional for a detector coupled to an operator of scaling dimension $\Delta$ and subjected to a Gaussian switching function exhibits several distinct features. First, the smooth Gaussian switching function eliminates the possible divergences associated with sudden switching. Second, this choice of switching function allows us to derive a closed-form expression for the decoherence rate. In long interaction time regime, the final result can be approximated to get the universal scaling law 
$\gamma \sim  a^{2\Delta-1}$. This result confirms that the scaling behavior is robust against the specific details of the switching protocol. It is also consistent with the scaling law previously derived using top-hat switching coupling.

In summary, the scaling exponent $(2\Delta-1)$ implies that the detectors coupling with the operators of higher scaling dimensions experience stronger Unruh-induced decoherence. For scalar operators ($\Delta=1$), the decoherence rate grows linearly with the proper acceleration, whereas for electromagnetic ($\Delta=2$) and fermionic ($\Delta=3$) fields, the decoherence rate scales as $a^3$ and $a^5$, respectively. This highlights the enhanced sensitivity of higher-dimension operators to probe the decoherence of the detector induced by the Unruh effect. Therefore, Gaussian switching can be viewed as a natural and physically meaningful protocol for studying Unruh-induced decoherence.

\section{Conclusion and discussion}

In this work we have investigated the decoherence of an Unruh-DeWitt detector coupled to scalar, electromagnetic, and fermionic fields along a uniformly accelerated trajectory. Using the influence functional formalism within the Schwinger-Keldysh framework, we derived the decoherence functional and established a universal scaling relation between the decoherence rate and the proper acceleration of the detector. The scaling law $\gamma \propto a^{2\Delta-1}$ reveals a clear hierarchy in the detector response determined by the scaling dimension $\Delta$ of the environmental field operator. As a consequence, the decoherence rate grows linearly with acceleration for scalar fields, cubically for electromagnetic fields, and as the fifth power of acceleration for fermionic fields.

We also examined the role of the detector switching protocol by comparing sharp (top-hat) and smooth Gaussian switching functions. While sharp switching provides a useful benchmark, Gaussian switching yields a well-controlled and physically realistic interaction profile, suppressing the possible divergency caused by the sudden coupling. In this case the decoherence functional admits a closed analytic expression. In the long interaction time limit, we have obtained the same universal scaling behavior for the decoherence rate.

Our results highlight that environments associated with higher-dimensional operators can significantly amplify Unruh-induced decoherence. This enhancement suggests that quantum detectors coupling to higher-dimensional operators may provide a more sensitive probe of acceleration-induced vacuum fluctuations than conventional scalar-field setups. Although direct observation of the Unruh effect through extreme mechanical accelerations remains challenging, the scaling behavior identified here may guide the design of analog quantum simulations or engineered quantum systems aimed at detecting acceleration-induced quantum noise. These findings open a possible avenue for exploring fundamental quantum field theoretic effects through decoherence dynamics in controllable quantum platforms. In summary, our work suggests that quantum decoherence may provide a promising route to Unruh effect beyond the conventional particle detection.

It should be noted that our derivation of the scaling law for the decoherence rate relies on the Lorentz-invariant, massless nature of the fields. As a potential future direction, it would be interesting to generalize the scaling law to the cases with the nontrivial dispersion relations, e.g., in Lorentz-violating setups \cite{Tian:2022gfa,Davies:2023zfq,Xu:2025smb}. In these cases, the form of the Wightman function and the corresponding decoherence rate of the detector would be modified by the generic dispersion relations, which could, in turn, affect the scaling behavior of the decoherence rate. Extending our analysis to such cases would be a valuable valuable direction to explore.

\acknowledgments

We thank Prof. Hongbao Zhang and Dr. Xidi Wang for useful discussions. R.L. would like to thank National Natural Science Foundation of China (No.12575059) and Shandong Provincial Natural Science Foundation (No.ZR2025MS10) for funding support. Z.X.M. would like to acknowledge support from the Shandong Provincial Natural Science Foundation under projects ZR2023LLZ015 and ZR2024LLZ012.

\appendix
\section{An alternative derivation of the decoherence functional}

In this appendix, we provide an alternative derivation of the decoherence functional in the scalar field environment. By expanding the scalar field in terms of the mode functions, it is shown that the decoherence dynamics can be governed by the displacement operators. This is consistent with the spin-boson model with a phase-damping coupling considered in \cite{Hornberger:2008xkz}. We than show that the relation between the expectation value of a product of displacement operators and the two point correlation function. 

We start with the interaction Hamiltonian given in Eq.\eqref{Int_Ham}. It is convenient to work in the interaction-picture. The unitary evolution operator is given by 
\begin{equation}
U_I = \mathcal{T} \exp\left[-i \int d\tau \; H_I(\tau)\right]=\mathcal{T} \exp\left[-i\lambda \int d\tau \;  \chi(\tau) \sigma_z \phi(x(\tau))\right] 
\end{equation}
For the scalar field operator $\phi(x)$, it can be expanded as 
\begin{equation}\label{scalar_exp}
\phi(x) = \sum_k \left(a_k u_k(t,\mathbf{x}) + a_k^\dagger u_k^*(t,\mathbf{x})\right),
\end{equation}
where $u_k(t,\mathbf{x})$ denotes the mode function and $a_k$ and $a_k^\dagger$ are the annihilation and creation operators. The index $k$ denotes a complete set of quantum numbers and the mode frequency. The annihilation and creation operators satisfy the conventional commutation relation 
\begin{eqnarray}
    [a_k,a_{k'}^\dagger]=\delta_{k,k'}.
\end{eqnarray}
The annihilation operator $a_k$ can be used to define the vacuum state as 
\begin{eqnarray}
    a_k |0\rangle =0.
\end{eqnarray}

By substituting the expansion into Eq.\eqref{Int_Ham}, it is easy to see that the interaction Hamiltonian has the similar form with the interaction Hamiltonian of the spin-boson model in \cite{Hornberger:2008xkz} 
\begin{eqnarray}
    H_I(\tau)=\lambda \chi(\tau) \sigma_z \sum_k \left(a_k u_k(x(\tau)) + a_k^\dagger u_k^*(x(\tau))\right)\;.
\end{eqnarray}
It is clear that the interaction is linear in the annihilation and creation operators. This makes the commutator of the interaction Hamiltonian at different time a $c$-number. Therefore, the unitary evolution operator can be exactly evaluated, which can be explicitly given by 
\begin{equation}
U_I = e^{i\Phi} \exp\left[-i \lambda \sigma_z \int d\tau\; \chi(\tau)  \sum_k \left(a_k u_k(x(\tau)) + a_k^\dagger u_k^*(x(\tau))\right)\right],
\end{equation}
where $\Phi$ is a real phase that is not relevant to the decoherence effect. By defining
\begin{equation}
\alpha_k \equiv -i \lambda \int d\tau \chi(\tau) u_k^*(x(\tau)),
\end{equation}
then the unitary evolution operator can be further written as 
\begin{equation}
U_I = \prod_k D_k(\sigma_z \alpha_k),
\end{equation}
where we have eliminated the irrelevant phase factor and defined the displacement operator as 
\begin{equation}
D_k(\alpha) = \exp\left( \alpha b_k^\dagger - \alpha^* b_k \right).
\end{equation}

As discussed in the main text, we consider the case that the initial state is a product state \eqref{Initial_state}, the reduced density matrix at late time can be given by 
\begin{equation}
\rho_d(t_f) = \mathrm{Tr}_\phi \left[ U_I \rho(t_i)U_I^\dagger \right].
\end{equation}
Using $\sigma_z |\pm\rangle = \pm |\pm\rangle$, one can find
\begin{equation}
U_I |\pm\rangle = \prod_k D_k(\pm \alpha_k) |\pm\rangle.
\end{equation}
Then the off-diagonal element of the reduced density matrix is given by 
\begin{equation}\label{rho_alpha}
\rho_d^{+-}(t_f)=\rho_d^{+-}(t_i)
\left\langle \prod_k D_k(2\alpha_k) \right\rangle_{\rho_\phi}=\rho_d^{+-}(t_i) \exp\left(-2\sum_k|\alpha_k|^2\right),
\end{equation}
where the initial state of the quantum field is set to be in the vacuum state. This equation indicates that the decoherence is determined by the expectation value of a product of displacement operators in the vacuum state of the environmental field. In addition, this derivation of the decoherence shows that the decoherence dynamics can be governed by the displacement operators. Physically, this derivation reflects that a classical source generates a classical field configuration, whose quantum counterpart is a coherent state.

Compared Eq.\eqref{rho_alpha} with Eq.\eqref{off_dia_rho}, the decoherence functional can be defined as 
\begin{eqnarray}
    \Gamma=2\sum_k|\alpha_k|^2=2\lambda^2 \int d\tau d\tau' \chi(\tau) \chi(\tau')  \sum_k u_k(x(\tau)) u_k^*(x(\tau'))  
\end{eqnarray}
Consider the Wightman function defined as 
\begin{eqnarray}
G_+(x,x') = \langle 0|\phi(x)\phi(x')|0\rangle \;.
\end{eqnarray}
By substituting the mode expansion of the scalar field operator into the above definition and using the definition of the vacuum state, one can obtain 
\begin{eqnarray}
   G_+(x,x') = \sum_k u_k(x) u_k^*(x'). 
\end{eqnarray}
Using this expression, the decoherence functional can be written as 
\begin{eqnarray}
   \Gamma=2\lambda^2 \int d\tau d\tau' \chi(\tau) \chi(\tau') G_H(x(\tau),x(\tau')) \;,
\end{eqnarray}
where we have used the symmetrized Hadamard function in the final expression.

In summary, we have derived the decoherence functional \eqref{Dec_GH} by using the method outlined in \cite{Hornberger:2008xkz}. Therefore, we have established that the influence functional obtained from the Schwinger-Keldysh path integral formalism is exactly equivalent to the expectation value of a product of displacement operators in the vacuum state of the environmental field. This also demonstrates the precise correspondence between our model and the phase-damping spin-boson model in \cite{Hornberger:2008xkz}.


 \bibliographystyle{JHEP}
 \bibliography{biblio.bib}

\end{document}